# Machine Learning of Energetic Material Properties

Brian C. Barnes[1], Daniel C. Elton[2], Zois Boukouvalas[2], DeCarlos E. Taylor[1], William D. Mattson[1], Mark D. Fuge[2], and Peter W. Chung[2]

[1]Energetic Materials Science Branch,
RDRL-WML-B, US Army Research Laboratory,
Aberdeen Proving Ground, MD, 21005-5066, USA

[2]Department of Mechanical Engineering, University of Maryland,
College Park, MD 20742, USA

**Abstract.** In this work, we discuss use of machine learning techniques for rapid prediction of detonation properties including explosive energy, detonation velocity, and detonation pressure. Further, analysis is applied to individual molecules in order to explore the contribution of bonding motifs to these properties. Feature descriptors evaluated include Morgan fingerprints, E-state vectors, a custom "sum over bonds" descriptor, and coulomb matrices. Algorithms discussed include kernel ridge regression, least absolute shrinkage and selection operator ("LASSO") regression, Gaussian process regression, and the multi-layer perceptron (a neural network). Effects of regularization, kernel selection, network parameters, and dimensionality reduction are discussed. We determine that even when using a small training set, non-linear regression methods may create models within a useful error tolerance for screening of materials.

## Introduction

Data-driven predictive modeling is revolutionizing fields as diverse as materials research[1-3], advertising, and control of autonomous vehicles.[4] This wide-ranging impact is driven by three main converging factors: cheaply available computing power, a rapid increase in the number and size of digitized datasets[5], and breakthroughs in "learning" algorithms for classification and regression problems[6-7]. In this work, we differentiate between "physics-based" and "data-driven" models. Physics-based models are built upon scientific theories that attempt to explain the problem being studied with an assumed causal relationship between inputs and outputs. The model physics may be extremely complicated and take hours (or days) to compute. Data-driven models apply a statistical model or learning algorithm in order to predict future observations with the best possible accuracy.[8-9] This predictive ability does not require any assumed causal relationship, and the predictive model may often be evaluated in a matter of seconds (or fractions thereof). The inputs and reference outputs from digitized datasets that are used for constructing a data-driven model may be obtained from experimental observation or extracted from another model of any type and fidelity. A data-driven model may still have some ability to provide physical insight into the nature of a process. In his article "To Explain or to Predict?"[9] Shmueli states that "Explanatory power and predictive accuracy are different qualities; a model will possess some level of each."[9]

For energetic materials, where experiments are inherently hazardous, a powerful, fast, and predictive model for arbitrary properties would reduce the number of required experiments to produce optimized results. In this work, we construct data-driven models using modern machine learning methods in order to make rapid predictions for detonation properties of energetic molecules.

**Methods**

Construction of a data-driven model requires selection of a useful set of input (data) with relevant information about the system expressed as an *N*-dimensional array. This is called a "featurization" or "descriptor". The data-driven model also needs a choice of underlying predictive algorithm and a determination of specific hyperparameters to be used for that algorithm.

### Feature Descriptors

A popular string representation for molecules is the Simplified Molecular-Input Line-Entry System (SMILES).[10] Working directly with molecular SMILES is not always convenient for regression tasks, as most regression algorithms require numerical arrays, instead of strings. A SMILES string may be transformed into a variety of other descriptors, or array representations.

The "sum over bonds" descriptor is a simple vector representation of bond types present in a molecule; a vector is constructed of length *N* where *N* is the number of unique bond types present in the set of molecules being studied. For each molecule, the descriptor contains an integer count of the number of times each type of bond is present.

The E-state vector is a physically motivated fixed-length fingerprint created to represent the "electrotopological state" of a molecule.[11] It was originally created for use with drug design studies.

The Morgan fingerprint, or extended-connectivity fingerprint, is another topological fingerprint and has user-controlled length.[12] It represents local connectivity over groups of atoms, and typically only represents the presence / lack of unique groupings through a binary representation.

A direct and global way to represent a molecule is by its "Coulomb matrix" representation, which by construction takes into account the 3D structure of the molecule.[13] For a given molecule, a Coulomb matrix requires a set of nuclear charges and the corresponding Cartesian coordinates of the atomic positions in a 3D space. By construction, the Coulomb matrix is invariant to translations and rotations of the molecule in the 3D space. However, they are not invariant under random permutations of the atom ordering. This issue can be resolved by using the eigenvalue spectrum of a Coulomb matrix as the molecule representation, since eigenvalues are invariant under permutation of rows or columns.

### Dimensionality Reduction

Representing the data using the eigenspectra of the Coulomb matrices is associated with a high-dimensional feature space. By transforming the data, we aim to obtain a set of features that encode the relevant information in a compact manner. Independent component analysis (ICA) is a popular approach to model reduction.[14] The basic noiseless ICA model is given by

$$\mathbf{x}(v) = \mathbf{A}\mathbf{s}(v), \quad v = 1, \ldots, V \quad (1)$$

where $v$ is the sample index, $\mathbf{s}(v) \in R^N$ are the unknown source signals, and $\mathbf{x}(v) \in R^M$ are the mixture data. A similar method is principal component analysis (PCA). Under the assumption that sources are statistically independent, the goal in ICA is to estimate a demixing matrix $\mathbf{W} \in \mathbb{R}^{N \times N}$ to yield maximally independent source estimates $\mathbf{y}(v) = \mathbf{W}\mathbf{x}(v)$.

### Regression algorithms

We will briefly discuss the four regression techniques used in this work with very brief descriptions. For a complete discussion of commonly used statistical learning methods, please see detailed references such as Hastie.[15]

### LASSO

Lasso regression performs a constrained minimization of the residual sum of squares of the difference between the predicted *($y^{pred}$)* and reference *($y^{ref}$)* values using the L1 norm of the regression coefficients **β** as a penalty:

$$min_{\beta,\lambda}\{\sum_i(y_i^{pred} - y_i^{ref})^2 - \lambda \sum_j|\beta_j|\} \quad (2)$$

Here, $\mathbf{y^{pred}} = \mathbf{\beta X}$, which is the product of the matrix of regression coefficients and descriptor array **X** across all items being evaluated. The L1 constraint acts a regularization term whose influence is controlled by the magnitude of the hyperparameter λ. For sufficiently large values of λ, some of the coefficients will approach a zero value resulting in a more parsimonious model that may be easier to interpret or more efficiently evaluated. Regularization is also used to prevent overfitting (as part of the well-known "bias-variance tradeoff").

Kernel Ridge Regression

Ridge regression shrinks regression coefficients by imposing a penalty on their size through the use of L2 regularization (a quadratic penalty on regression coefficients). The "kernel trick" is used when making a prediction $\mathbf{y}'$, given by

$$\mathbf{y}' = \mathbf{y}^\mathrm{T}(\mathbf{K} + \lambda\mathbf{I})^{-1}\kappa \qquad (3)$$

where $\mathbf{K}$ denotes the empirical kernel matrix, $\mathbf{I}$ the identity matrix, $\mathbf{y}$ is the vector of reference values, and $\kappa$ is a product of reference and target descriptors. The "kernel trick" allows for efficient evaluation of the model in a high dimensional space.

Gaussian Process Regression

A Gaussian process is defined by a collection of random variables [$\mathbf{x}$] where any finite subset of the variables has a joint distribution of Gaussian form. The Gaussian process is specified in a manner similar (but not identical to) that of kernel ridge regression. In this work, the rational quadratic kernel was used as the covariance function for Gaussian process regression. Gaussian process regression, being based in a Bayesian formulation and assuming a prior distribution regarding noise in the data, is able to provide predictions not just for the mean value of observations, but also for the variance (or uncertainty) in individual predictions.

Neural Network

We use a fully-connected feed-forward network, the multi-layer perceptron. Neural networks are powerful tools for both classification and regression problems. The multi-layer perceptron is composed of multiple layers, each layer is composed of nodes, and all nodes in adjacent layers are connected by activation functions. Each node in the input layer represents a feature in an input descriptor. The number of hidden layers and nodes per layer are optimized as hyperparameters. There is a single output node, which holds the numerical output of the regression model. In this work we use hyperbolic tangent activation functions for the hidden layers and linear activation for the output layer. Data is scaled to zero mean and unit variance before training, and transformed to original scale upon output. Training of the neural network is performed using the "Adam" stochastic optimization algorithm. L2 regularization is applied to activation weights.

Data

The featurization and regression algorithms described above are applied to a dataset sourced from two open literature articles reporting predicted detonation properties and other thermodynamic quantities for a total of 416 molecules.[16-17] The properties in this dataset are a mix of results from thermochemical codes and empirical Kamlet-Jacobs relations. It is well-known that a thermochemical code and the Kamlet-Jacobs equations will not give the same result for prediction of detonation properties. This may be viewed as similar to adding noise to the inputs, or as similar to sourcing experimental data from labs with different testing conditions. This also provides information about scaling of algorithm performance with training set size.

**Results**

Models for Detonation Pressure and Velocity

We have constructed models to predict detonation pressure and detonation velocity of energetic molecules. Given the skeletal formula for a molecule (in the form of a SMILES string), these models return a result in less than one second. The model hyperparameters were optimized using a simple grid search, and error metrics were evaluated using nested 5-fold cross-validation.[18-19] In this section, we provide models using LASSO, Gaussian process regression, and neural network algorithms. Results are provided in table 1.

|  | MAE | RMSE | $Q^2$ | Pearson |
|---|---|---|---|---|
| **Detonation Velocity** | | | | |
| *LASSO* | | | | |
| SoB | 0.269 | 0.369 | 0.74 | 0.859 |
| E-state | 0.273 | 0.371 | 0.74 | 0.859 |
| Morgan | 0.327 | 0.441 | 0.63 | 0.796 |
| *GPR* | | | | |
| SoB | 0.163 | 0.281 | 0.85 | 0.921 |
| E-state | 0.181 | 0.299 | 0.83 | 0.911 |
| Morgan | 0.311 | 0.424 | 0.66 | 0.812 |
| *Neural Net* | | | | |
| SoB | **0.159** | **0.267** | **0.86** | **0.931** |
| E-state | 0.173 | 0.278 | 0.85 | 0.924 |
| Morgan | 0.333 | 0.453 | 0.61 | 0.796 |
| **Detonation Pressure** | | | | |
| *LASSO* | | | | |
| SoB | 2.03 | 2.78 | 0.78 | 0.886 |
| E-state | 2.09 | 2.85 | 0.77 | 0.880 |
| Morgan | 2.80 | 3.83 | 0.59 | 0.774 |
| *GPR* | | | | |
| SoB | 1.36 | 2.37 | 0.85 | 0.927 |
| E-state | 1.51 | 2.40 | 0.84 | 0.916 |
| Morgan | 2.54 | 3.51 | 0.66 | 0.814 |
| *Neural Net* | | | | |
| SoB | **1.32** | 2.19 | 0.86 | 0.932 |
| E-state | 1.33 | **2.15** | **0.87** | **0.934** |
| Morgan | 2.86 | 3.93 | 0.57 | 0.776 |

Table 1. Mean absolute error (MAE), root-mean-squared error (RMSE), coefficient of determination on the test set ($Q^2$), and Pearson correlation coefficient; all scoring metrics are for the test set. Models reported include LASSO, Gaussian process regression (GPR), and a neural network, each with three descriptor sets: sum over bonds (SoB), the E-state fingerprint, and Morgan fingerprint. Best performing results are highlighted in bold.

Nested cross-validation (used for results in table 1) is set up as a nested loop. The outer loop partitions the data $N$ times, with each observation appearing in one of the $N$ test sets exactly once. The holdout data in those $N$ test sets is used for calculation model errors, which is then averaged across the $N$ different models (each evaluated on a unique holdout set) to provide the error metrics in table 1. That is referred to as the "outer loop". For the inner loop, the non-holdout data for each of the $N$ folds is used in k-fold cross-validation for hyperparameter optimization, with the resulting best hyperparameters fit against the entire set of non-holdout data for that $N$th fold. Therefore, nested CV leverages cross-validation for both parameter selection and model evaluation, and avoids an "optimistic" selection bias that may result from using k-fold cross-validation alone. In this work, we used 5 folds for each nested loop.

As a result of using nested CV, we are also able to calculate standard deviation for the outer loop (model selection) scores. This serves as uncertainty quantification for our model selection and provides a meaure of the robustness of the hyperparameter solutions. For prediction of detonation pressure using the neural net, the SoB and E-State descriptors have similar performance. Although the SoB descriptor has a lower MAE, the standard deviation on MAE across the $N$ outer folds is 0.22 GPa for the SoB descriptor and 0.16 GPa for the E-state descriptor, placing model performance well within error of each other. This is the case for all four error metrics for those models. Gaussian process regression using the SoB descriptor is also within the nested CV error estimates for the best neural network methods. We note that the Morgan fingerprint performance is still worse than the other descriptors after accounting for a 95% confidence interval. As the Morgan fingerprint itself has hyperparameters, which we did not attempt to optimize, a future effort might improve the results using that descriptor.

The mean absolute error in detonation velocity for the best regularized linear model is nearly 70% higher than the error in the neural network model. In figures 1 and 2, we show plots of predictions versus reference values for detonation velocity and pressure, using neural network models. The error metrics reported in figures 1 and 2 are calculated for the entire set of outer loop holdout data; this is in contrast to table 1, which reports averages of the outer folds. This results in a slight difference in RMSE and Pearson correlation. Examination of figures 1 and 2 shows a relatively consistent behavior for predictions of either property, with the greatest error being for the highest performing molecules. This is somewhat counter-intuitive as the training set contains many energetic molecules, but it is likely due to the largely local (nearest-neighbor) nature of descriptors, which would not account for factors such as ring strain or structures leading to chain reactions. Future work will seek to improve on accuracy for these molecules.

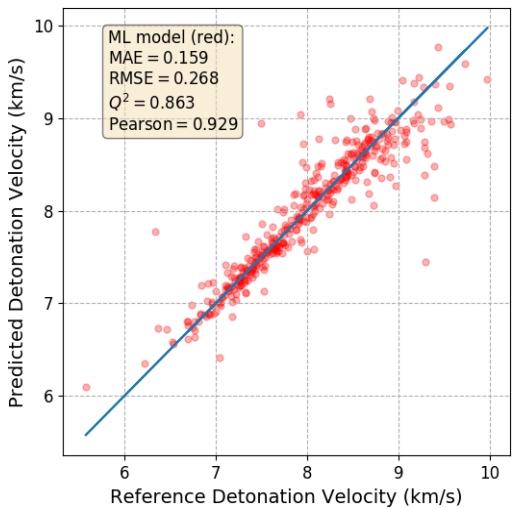

Figure 1. Predicted detonation velocity (km/s) versus reference detonation velocity, for nested cross-validation test set samples, and error metrics.

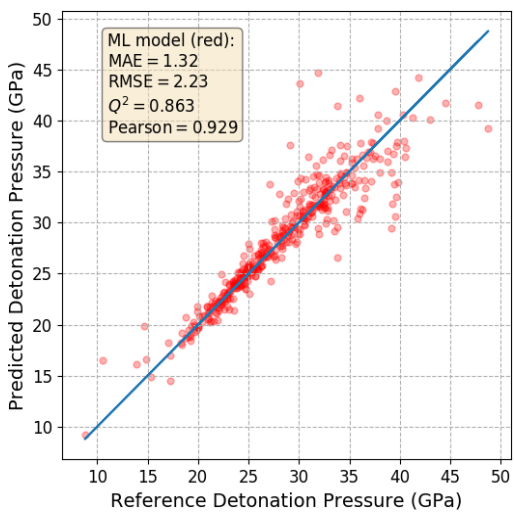

Figure 2. Predicted detonation pressure (GPa) versus reference detonation pressure, for nested cross-validation test set samples, and error metrics.

The overall low performance of the Morgan fingerprint is likely due, in part, to its default representation being a binary measure of the presence of localized functional groups, instead of an extensive measure. For example, the Morgan fingerprint bit vector (1024 bits) with radius 2 is identical for RDX and HMX. A more detailed investigation of descriptor performance is available elsewhere.[20]

Critical to constructing a strong performing neural network were the choices of activation function (hyperbolic tangent) and scaling of data to zero mean and unit variance before training the network. Future neural network models will evaluate additional activation functions and network optimization strategies, as well as network topologies designed for use with molecular graphs. The use of the rational quadratic kernel allowed for Gaussian process regression to outperform the reference LASSO model. A benefit of using a Gaussian process model is that predictions of mean values may be made with an accompanying prediction of variance, reflecting uncertainty in the model for any given prediction. While this is possible in some neural network models using dropout techniques, it is not commonly applied and still a topic of active research.[21]

Dimensionality Reduction with KRR model

To demonstrate the effect of PCA and ICA on the regression results when using Coulomb matrices, we perform the following experiment. By using the eigenspectrum of all 416 molecules we perform PCA and then ICA on the entire dataset $\mathbf{X} \in \mathbb{R}^{87 \times 416}$, where 87 is the dimension of each feature vector. The transformed dataset is denoted by $\hat{\mathbf{X}} \in \mathbb{R}^{10 \times 416}$ and as it can be observed that the dimension of each feature vector has been reduced to 10. Additionally, due to the effect of ICA, we can assume that the estimated sources in $\hat{\mathbf{X}}$ are as independent as possible. Table 2 shows the training and testing MAE of detonation velocity using 5-fold kernel ridge regression on $\mathbf{X}$ and $\hat{\mathbf{X}}$. Results reveal that PCA and ICA can be used to obtain low dimensional feature vectors that reduce the test error significantly. The test error for the relatively small set of 416 molecules benefited from the dimension reduction by approximately 5%.

| 5-fold | MAE | |
|---|---|---|
| KRR | Train | Test |
| $\mathbf{X} \in \mathbb{R}^{87 \times 416}$ | 0.296 | 0.400 |
| $\hat{\mathbf{X}} \in \mathbb{R}^{10 \times 416}$ | **0.289** | **0.377** |

Table 2. Mean absolute error for the prediction of detonation velocity using the original and the transformed datasets.

Interpretation of Molecular Features

Interpretation, broadly construed, is being able to explain what a model is doing when making a prediction in terms that a subject matter expert would understand. In the context of machine learning applied to energetic molecules, there are many uses for interpretation:
- To ensure the featurization and model is capturing known structure-property relationships. If it is not, this may suggest ways to improve the featurization and model.
- To discover structure-property relations the model is using which conflict with physical theory (this can be due to biased training data, overfitting, or even new physics).
- To discover latent variables (combinations of features) the model is using that may be useful for molecular designers.

In this report, we will focus on two general approaches: sensitivity analysis of models and feature importance ranking techniques.

In standard sensitivity analysis, each feature in the feature vector is changed by a small amount across the dataset while holding the others constant, and the change in the model's accuracy is recorded. The type of sensitivity analysis we study is the "similarity map" scheme of Riniker and Landrum.[22] This approach differs from conventional sensitivity analysis by removing individual atoms in each molecule from consideration in the featurization and quantifying how this affects the model's output.[22] The method is implemented in *RDKit*, including a "topo-map" visualization scheme which places a Gaussian peak on each atom with fixed width but a variable height corresponding to how sensitive the model is to the presence of that atom. The similarity map technique has been used to help molecular designers find molecules with reduced skin sensitivity[23] and cardiac toxicity.[23]

We tested the similarity maps technique using the same dataset as above. We trained a kernel ridge model using three featurizations for which we have atom-level sensitivity analysis implemented: the *RDKit* "custom" fingerprint, the Atom-Pair fingerprint, and our combined featurization. The combined featurization combines the E-state fingerprint descriptors, the sum over bonds featurization, and our custom descriptor set which is described elsewhere.[20]

Figure 3 shows the resulting "heat maps" for three representative molecules: RDX, CL-14, and CL-20. Despite the models having similar average test errors, the sensitivity analyses for these models are quite different. After studying several dozen visualizations (not shown) for each featurization it was hard to pick out general trends, with the exception of the *RDKit* custom fingerprint. In the *RDKit* fingerprint the visualizations aligned well with chemical intuition; nitrogen atoms appeared green or white and hydrogen atoms appeared red almost uniformly. We hypothesize this is due to the fact that this featurization explicitly counts common fragments (small bonding arrangements of 2-5 atoms, such as functional groups). Functional groups play an important role in the chemical theory behind energetic properties and also the way people think about molecules. The lack of consistent results from the other featurizations suggests that using this method for discovering chemical insights should be used with great care. However, despite the negative result presented here we believe this avenue of interpretation is worth exploring further; larger and more diverse datasets could yield more consistent results and even new chemical insights. Incorporating additional non-energetic molecules into the data should also improve the results, since models would be forced to distinguish important features which make molecules energetic.

Feature Ranking

Within QSPR/QSAR, feature ranking is employed routinely to do dimensionality reduction, interpret model behavior, and illuminate structure-property associations in data. In table 3 we compare several of the most popular feature ranking methods. The simplest method of feature importance ranking is to use the magnitude of Pearson correlation. In our rankings, we only include features whose correlation is below the $p < 0.01$ level. That is, we require that the null hypothesis (no correlation) is rejected with a probability of less than 1% that the result comes from random (Gaussian) fluctuations. While $p$-values should be used with extreme care[24], they are useful for exploratory analysis. Some advantages of Pearson correlation are that it is easy to understand and that both positive and negative associations are delineated. The disadvantage of Pearson correlation is that it is only sensitive to linear dependencies. The mutual information (MI) correlation coefficient,[25] and maximal information criteria (MIC),[26] by contrast are sensitive to nonlinear dependencies. Next, we looked at using LASSO (described above) for feature ranking. The absolute magnitude of coefficient size in LASSO regression is often used as a means of ranking feature importance.

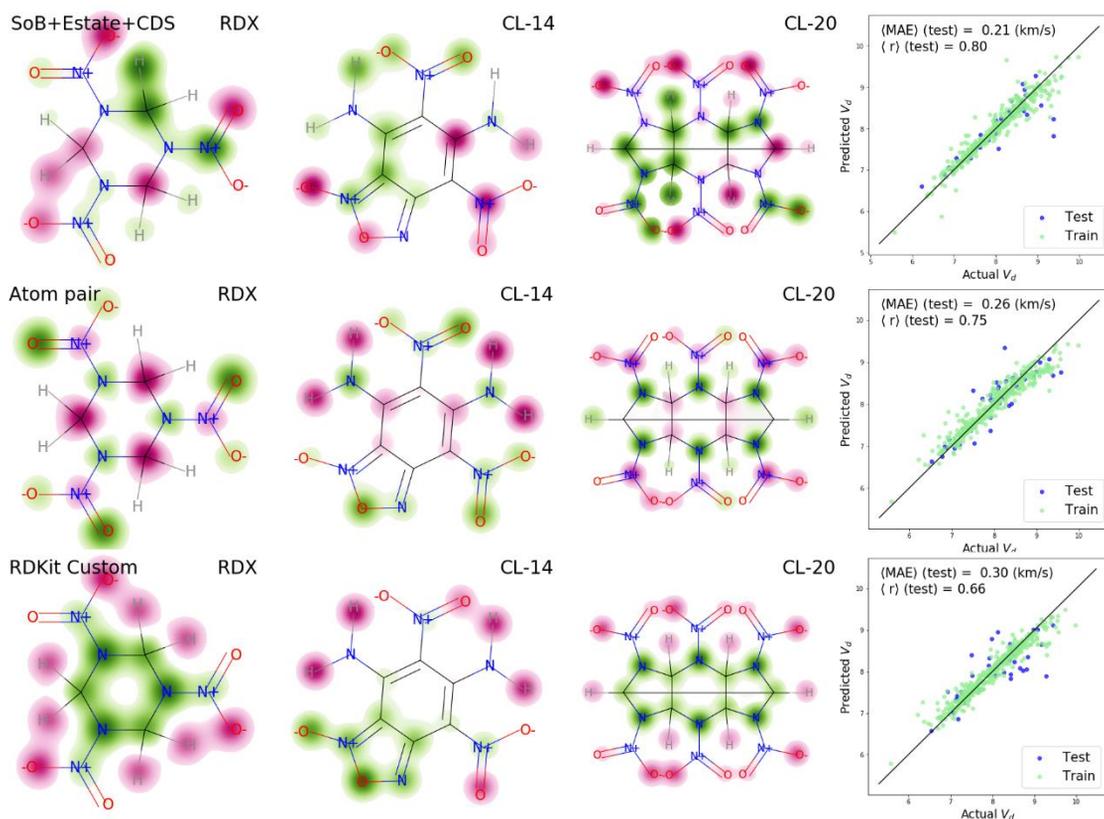

Figure 3. Comparison of our atom-level sensitivity analysis for the prediction of detonation velocity. Three representative molecules are shown (RDX, CL-14, and CL-20). Gaussian peaks of equal width but varying height are placed on each atom in accordance to that atom's effect on the models prediction. Red (negative) peaks indicate atoms whose removal increased the model's prediction, indicating they are associated with decreased detonation velocity. Green (positive) peaks indicate atoms whose removal decreased the model's prediction, indicating they are associated with increased detonation velocity. The visual clarity of the diagram was improved by cutting off colorization below a lower threshold. Diagrams indicating the overall accuracy of the models in the train and test data are shown in the right hand column. The mean absolute error (MAE) and Pearson correlation ($r$) values are for the test data, averaged over 5 random train-test splits.

Tuning the LASSO regularization strength parameter α by hand was important to obtain informative rankings. The aforementioned techniques (Pearson, MI, MIC, and LASSO) are all examples of the univariate feature ranking method because they treat one feature at a time. Two non-informative features can be informative when combined, however, so multivariate methods are also worth exploring. The technique of stability selection[27] is a multivariate method which studies the performance of a model across many different subsets of features and selects features which have high value over many subsets. Here, stability selection was implemented with LASSO regression. Finally, we looked at two ways of interpreting a random forest model (an ensemble of decision trees) to do feature importance ranking: scrambling (mentioned earlier) and variance score ranking. In variance score ranking, features which on average appear higher in decision trees are considered more important. In contrast to the results from atom level sensitivity analysis, these results are much more consistent. All of the methods rank two features, oxygen balance and aromatic carbons (designated as *aCa* or *C:C*), within the top two features. Oxygen balance is a widely used heuristic, and aromatic carbons indicate an aromatic ring, which the results indicate is associated with lower detonation velocity. C-H and N-N bonds are associated with lower and higher detonation velocity, respectively, in accordance with chemical intuition. Nitro groups bonded to nitrogen were consistently ranked higher than nitro groups attached to carbon. We believe such feature rankings, especially when analyzed over several target properties (pressure, sensitivity, density, etc.) could be useful to designers in the future; for now, these results show that these types of cheminformatics analyses are consistent with chemical intuition for energetic molecules.

Caveats and pitfalls

Model interpretation and feature ranking results can serve as a complement to established chemical heuristics and physics based techniques, but many caveats must be borne in mind when attempting to draw conclusions from such methods. First, machine learning methods primarily study correlation, and the presence of causation behind a correlation is never guaranteed. The presence of a correlation of X with Y may be due to the presence of a confounding variable, Z, where Z is causing the appearance of both X and Y. However, there are additional techniques which can help isolate causal mechanisms. Matched molecular pairs analysis (MMPA) studies molecules which differ by the addition or removal of a single functional group, yielding insights into the causal effect of such groups on target properties.[28-29] A small or non-diverse dataset can introduce significant sampling bias, and spurious correlations. Spurious correlations often occur when the number of data points is close to the number of features used.[30-32] The probability of spurious correlations can be rigorously quantified using false discovery rate techniques. We do not believe spurious correlations play a significant role our feature rankings, but we hypothesize they do play a role in the sensitivity maps. Further work is being done to rigorously quantify the probability of spurious correlations given the dataset used in this study.

**Conclusion**

We have demonstrated the successful application of data-driven, machine learning techniques for prediction of detonation properties and analysis of molecular-level features. The methods used closely follow recent approaches adopted by the pharmaceutical industry for computational drug design studies. The primary result in this work is the successful optimization of a neural network architecture and a Gaussian process model to provide results of greater accuracy than other (linear and nonlinear) methods with regularization, while using simple, readily-available descriptors. This provides a path forward for high-throughput screening of candidate molecules using very high-accuracy methods and with uncertainty quantification for individual predictions. Ongoing and future work involves the creation and curation of at least three new datasets: a much larger training set (including many non-energetic molecules) derived from quantum mechanics and a thermochemical code, a training set derived from experimental data targeting properties such as thermal decomposition of energetics, and a training set focused on energetic formulations. With new datasets in hand, we may construct models of much higher quality and with some immediate practical applications. Future work also involves creation of generative neural networks for suggestion of novel high-performance molecules, and extension of neural network techniques to predicting synthetic pathways for energetic material precursors.

| ranking | Pearson correlation | | Mutual information | | Maximal information criteria | | LASSO coefficient size | | LASSO stability selection | | Random forest variance score | | Random forest shuffling | |
|---|---|---|---|---|---|---|---|---|---|---|---|---|---|---|
| 1 | $OB_{100}$ | +0.760 | $OB_{100}$ | 0.730 | $OB_{100}$ | 0.654 | $OB_{100}$ | +0.490 | $OB_{100}$ | 1.000 | $OB_{100}$ | 0.692 | $OB_{100}$ | 1.131 |
| 2 | C:C | -0.498 | C:C | 0.267 | C:C | 0.352 | C-H | -0.359 | C-H | 1.000 | C-H | 0.045 | C:C | 1.011 |
| 3 | aCa | -0.430 | C-H | 0.211 | aCa | 0.307 | C-C | +0.268 | C-C | 0.950 | >C< | 0.039 | $n_{CNO_2}$ | 1.008 |
| 4 | N-N | +0.405 | aCa | 0.211 | N-N | 0.274 | $n_CO$ | -0.266 | =N | 0.825 | =O | 0.030 | aCa | 1.005 |
| 5 | $n_{NNO_2}$ | +0.384 | =O | 0.175 | >C< | 0.263 | H-O | +0.246 | N-N | 0.685 | C-O | 0.022 | N=O | 1.003 |

Table 3. Comparison of different feature ranking techniques, showing the top five features which are associated with detonation velocity across the dataset. ":" or "a" denotes an aromatic bond, "-" a single bond, and "=" a double bond. $OB_{100}$ is oxygen balance, and $n_{NNO2}$ is the number of nitro groups bonded to a nitrogen. Other features here are the Estate fingerprint descriptors representing atom types with different sigma and pi bonding configurations.[11]